\documentclass[a4paper]{article}
\usepackage[utf8]{inputenc}
\usepackage{anysize}
\usepackage{graphicx}
\usepackage{mathtools}
\usepackage{amsmath}
\usepackage{bm}
\usepackage{enumerate}
\usepackage{multirow}
\usepackage{algpseudocode}
\usepackage{algorithmicx}
\usepackage{subfig}
\usepackage{caption}
\marginsize{2.5 cm}{2.5 cm}{2.5 cm}{2.5 cm}

\title{\textbf{A New Algorithm for Updating and Querying Sub-arrays of Multidimensional Arrays}}
\author{Pushkar Mishra\\
Computer Laboratory,\\
University of Cambridge\\
\texttt{pm576@cam.ac.uk}}
\date{\vspace{-7 mm}}
\begin{document}

\maketitle
\begin{abstract}
Given a $d$-dimensional array $A$, an \textit{update} operation adds a given constant $C$ to each element within a continuous sub-array of $A$. A \textit{query} operation computes the sum of all the elements within a continuous sub-array of $A$. The one-dimensional update and query handling problem has been studied intensively and is usually solved using segment trees with \textit{lazy propagation} technique. In this paper, we present a new algorithm incorporating Binary Indexed Trees and Inclusion-Exclusion Principle to accomplish the same task. We extend the algorithm to update and query sub-matrices of matrices (two-dimensional array). Finally, we propose a general form of the algorithm for $d$-dimensions which achieves $\mathcal{O}(4^d*\log^{d}n)$ time complexity for both updates and queries. This is an improvement over the previously known algorithms which utilize hierarchical data structures like quadtrees and octrees and have a worst-case time complexity of $\Omega(n^{d-1})$ per update/query.
\end{abstract}


\section{Introduction}
The problem of updating and querying sub-arrays of multidimensional arrays is of consequence to several fields including data management, image processing and geographical information systems. The one-dimensional version of this problem has conventionally been solved using segment trees with lazy propagation technique. We show in this paper that a $d$-dimensional segment tree ($d > 1$) supports lazy propagation only along one out of the $d$ dimensions. Consequently, the worst-case time complexity for updates and queries becomes $\mathcal{O}(n^{d-1}\log n)$ instead of $\mathcal{O}(\log^{d}n)$.

\vspace{2 mm}
Space-partitioning hierarchical data-structures like the quadtree --- and its generalization to higher dimensions, \textit{e.g.}, octree --- perform better than multidimensional segment trees.\cite{HS1} These structures work by recursively dividing the given $d$-dimensional space ($d > 1$) into smaller axis-parallel hyper-rectangles. Such trees have a worst-case time complexity of $\Omega(n^{d-1})$ for updates and queries.

\vspace{2 mm}
The algorithm that we propose is based on Binary Indexed Trees\cite{BIT} (BITs) and inclusion-exclusion principle. It has the same space complexity as the aforementioned algorithms; but the worst-case time complexity for updates and queries is $\mathcal{O}(4^d*\log^{d}n)$.

\vspace{2 mm}
\noindent
For the purpose of this paper, we define three sets of operations.

\begin{enumerate}
\item \textit{Point-update range-query}\\
Point-update range-query refers to the set of operations wherein updates are performed only on unit-sized sub-arrays (\textit{i.e.}, individual elements) while queries are performed on sub-arrays of arbitrary sizes.
\item \textit{Range-update point-query}\\
Range-update point-query refers to the set of operations wherein updates are performed on sub-arrays of arbitrary sizes while queries are performed only on unit-sized sub-arrays.
\item \textit{Range-update range-query}\\
Range-update range-query refers to the set of operations wherein both updates and queries are performed on sub-arrays of arbitrary sizes. Note that the first two sets of operations are subsets of this set.
\end{enumerate}

\noindent
\textbf{Problem Definition.} Given is a $d$-dimensional array $A$ of size $N = n_1\times n_2 \times \dots \times n_d$, where $n_k$ ($1 \leq k \leq d$) is the length of the $k^{th}$ dimension. For the sake of simplicity, we assume that $n_1, n_2, \dots, n_d = n$. The goal of the problem is to handle on-line range-update range-query operations on $A$. By on-line we imply that a particular operation must be performed before the next one can be handled, and there is no prior knowledge of the order of operations.

\vspace{2 mm}
Using the known algorithms for the first two sets of operations, we devise an efficient algorithm for the aforementioned problem. Note that throughout this paper, we assume a unit-cost word Random Access Machine (RAM) model with word size $\Theta(\log n)$. On such a model, standard arithmetic and boolean bitwise operations on word-sized operands can be performed in $\Theta(1)$ time.

\vspace{2 mm}
\noindent
\textbf{Paper organization.} In section 2, we discuss the motivation behind this work. In Section 3, we summarize our contributions. In section 4, we define the terms and notations used in this paper. We give the formal problem statement in section 5. In section 6, we discuss the existing algorithms for handling \textit{range-update range-query} operations on multidimensional arrays. In section 7, we provide a concise description of Binary Indexed Trees and associated algorithms. We describe the new algorithm and its time and memory complexities in section 8. In section 9, we provide the data obtained from experimental comparison between execution times of the old and new algorithms. Lastly, we offer brief conclusions in section 10.


\section{Background and motivation}
Representation of a $d$-dimensional space or object as a $d$-dimensional array has numerous applications in several fields of Computer Science. For example, images can be represented as 2D arrays where individual cells can represent pixels. Geographical regions can also be represented as 2D arrays with individual cells denoting unit area. Space-partitioning, hierarchical data structures like the quadtree (and its generalization to higher dimensions, e.g., octree) are used perform a variety of operations on such representation of space and objects\cite{HS2}.

\vspace{2 mm}
Motivation for this paper stems from the need to process \textit{range-update range-query} operations with better time complexity. Geographical Information Systems(GIS) are an example where faster processing can be of significance.  In their paper\cite{GIS}, Samet \textit{et al.} describe the development of a GIS which uses quadtree as the underlying data structure. Consider the case when different regions of a country have received different amounts of rainfall, and total rainfall needs to be calculated for the season. Consider another case when different regions in a country have received different amounts of water in supply for a month, and the net water supplied is to be calculated for the country. These are cases of Range-update and \textit{Range-query} in a GIS respectively. Directly visiting each unit cell of a particular region to add or query certain a parameter can prove to be time taking, specially if the unit cells represent very small areas, i.e., the GIS is detailed. Therefore, Samet uses a quadtree for processing updates and queries. We show in this paper that these types of operations can be performed with better time complexity using Binary Indexed Trees.

\vspace{2 mm}
The need for processing updates and queries is not limited to one or two dimensions. Thus, it is important to have an algorithm which can be extended to any number of dimensions, without much effort.


\section{Our Contributions}
The main contributions of this paper are as follows.

\begin{enumerate}
\item Discussion and analysis of known algorithms.
\item A new algorithm for processing on-line \textit{range-update range-query} operations on multidimensional arrays.
\item Generalization of the algorithm to arbitrary number of dimensions.
\item Experimental comparison between execution times of the old and new algorithms for two and three dimensional arrays.
\end{enumerate}


\section{Preliminaries}
Let us first provide some general definitions for clarity. By $\log n$, we mean $\log_2 n$. An array refers to a 1D array unless stated otherwise. Both sub-array and `range' refer to a contiguous portion of an array.

\vspace{2 mm}
A sub-array of a $d$-dimensional array $A$ is depicted as $A[a_1, b_1, c_1, d_1, ... :\, a_2, b_2, c_2, d_2, ...]$, where $a_1$, $a_2$ are coordinates along the $1^{st}$ dimension, $b_1$, $b_2$ are coordinates along the $2^{nd}$ dimension, and so on. It consists of all the elements $arr[a][b][c][d][...]$ such that $a_1 \leq a \leq a_2$, $b_1 \leq b \leq b_2$, $c_1 \leq c \leq c_2$, $d_1 \leq d \leq d_2$, and so on.\\
A general sub-array is denoted by $[a_1, b_1, c_1, d_1, ... :\, a_2, b_2, c_2, d_2, ...]$.

\vspace{2 mm}
Throughout this paper, we use $rsum(a_1, b_1, ... :\, a_2, b_2, ...)$ to refer to the cumulative sum of all the elements of the sub-array $[a_1, b_1, ... :\, a_2, b_2, ...]$.

\vspace{2mm}
We denote a point in $d$-dimensional space by $(x_1, x_2, x_3, \dots, x_d)$. A point in 1D space is denoted without parentheses.

\vspace{2 mm}
$bitAnd(a, \, b)$ refers to the bitwise and of integers $a$ and $b$. For example, $bitwise(2,\, 3)$ will return 2 since 2 and 3 are represented as $10$ and $11$ in binary.

\vspace{2 mm}
We recommend that the reader be familiar with Binary Indexed Trees and basic operations on them\cite{BIT} in order to understand the algorithms and concepts being put forward.


\section{Problem statement}
In this section, we put forward the formal problem statement.

\vspace{2 mm}
Given is a 1D array $A$ of length $n$ with all the elements initially set to 0. Two types of operations need to be performed on this array:

\begin{enumerate}
\item Given $x_1$ and $x_2$, add a constant $c$ to all the elements of the sub-array $A[x_1 : x_2]$.
\item Given $x_1$ and $x_2$, output the sum of all the elements of the sub-array $A[x_1\, :\, x_2]$.
\end{enumerate}
We refer to this as the `1D version' of the problem.

\vspace{2 mm}
The problem can be extended to any number of dimensions. Consider the 2D version. Given is a matrix $M$ with side-length $n$ with all the elements initially set to 0. Again, two types of operations need to be performed on this matrix:
\begin{enumerate}
\item Given $(x_1$, $y_1)$ and $(x_2$, $y_2)$, add a constant $c$ to all the elements of the sub-matrix $M[x_1, y_1\, : \,x_2, y_2]$.
\item Given $(x_1$, $y_1)$ and $(x_2$, $y_2)$, output the sum of all the elements of the sub-matrix $M[x_1, y_1 \,: \,x_2, y_2]$.
\end{enumerate}

\vspace{2 mm}
\noindent
In this paper, we describe our algorithm for 1D and 2D versions of the problem, and then generalize the idea to $d$-dimensions.


\section{Previous work}
In this section of the paper, we discuss the previously known algorithms and data structures for handling on-line range-update range-query operations. We analyze their time and space complexities.


\subsection{For 1D version}
1D version of the problem has conventionally been solved using segment trees. De Berg \textit{et al.}, in their book \cite[p. 226]{ST}, prove that any interval $[i : j]$ can be constructed by using a maximum of $\mathcal{O}(\log N)$ nodes of the segment tree. The lazy propagation technique ensures that no more than $\mathcal{O}(\log n)$ nodes are to be visited to perform the required update or query. In this technique, the node in the segment tree, which contains the range to be updated, is marked (or `flagged'), and the update is propagated down to its children only when a query is to be performed on that node, or on its children.

\vspace{2 mm}
Since, a maximum of $\mathcal{O}(\log n)$ nodes must be visited for performing update or query, the running time of this algorithm is $\mathcal{O}(\log n)$ per update/query operation. The space required to execute it is $\mathcal{O}(n)$.\cite[p. 226-227]{ST}


\subsection{For Higher Dimensions}
The segment tree can be generalized to any number of dimensions in the form of multilevel segment trees. For example, for a $d$-dimensional array with side-length $n$, we first build one-dimensional segment trees along the $d^{th}$ dimension (also referred to as `last dimension'). We then build segment trees along $(d-1)^{th}$ dimension, and so on. 

\vspace{2 mm}
In multidimensional versions, the segment tree stores a collection of axis-parallel hyper-rectangles. \textit{Point-update range-query} or \textit{range-update point-query} operations can be implemented in $\mathcal{O}(\log^d n)$ complexity. However, we show that a multidimensional segment tree does not produce an optimal solution in the case of \textit{range-update range-query} operations.

\vspace{2 mm}
\noindent
\textbf{Proposition 1.} A $d$-dimensional segment tree does not support Lazy Propagation technique on more than one of the dimensions.

\vspace{1 mm}
\noindent
\textit{Proof.}
We consider a 2D segment tree, built on a matrix $mat$. As described earlier, in building the 2D segment tree, we first build segment trees along each row, and then along columns. Each of the node in this 2D segment tree stores sum of elements of a sub-matrix of $mat$. From the manner in which it is build, every individual element $(x, y)$ of $mat$ is contained in $\mathcal{O}(\log^2 n)$ nodes of the 2D segment tree. Of these $\mathcal{O}(\log^2 n)$, we take two nodes $l$ and $r$. We choose $l$ and $r$ such that the sub-matrix contained by $l$ does not completely lie inside the sub-matrix contained by $r$ and vice versa. We now perform a lazy-update on node $l$. Since $l$ and $r$ have at least one common element, the value contained at $r$ is also affected by the update. But, when value at $r$ is queried, it would not return the updated value. This is because, in a segment tree, a query always moves towards the root. The path from node $r$ to the root of the segment tree would not contain $l$, since the sub-matrix associated with $l$ does not fully contain the sub-matrix associated with $r$. Consequently, the update `flag' at $l$ is never encountered during query on $r$. Therefore, Lazy Propagation technique does not produce the correct result.

\vspace{2 mm}
As a result of \textit{Proposition 1}, the running time for each operation (update or query) on a $d$-dimensional segment tree becomes $\mathcal{O}(n^{d-1}\log n)$. This is because, for each of the $d-1$ dimensions that does not support lazy propagation, we need to recursively visit individual indexes in the range to be updated or queried. As proved by \textit{Berg et. al}, a segment tree possesses $2n$ nodes \cite[p. 226-227]{ST}. Consequently, memory requirement of a $d$-dimensional segment tree is given by $\mathcal{O}((2n)^d)$.

\vspace{2 mm}
As stated in the first section, quadtrees and their generalization to higher dimensions perform better than multidimensional segment trees. Below, we analyze their time and space complexity.

\vspace{2 mm}
A quadtree\cite{QT1} recursively divides the given 2D space into four quadrants. Quadtrees can be used to perform \textit{range-update range-query} operations on matrices in the same manner as segment trees can be used for 1D arrays. For simplicity, let us assume that the side-length of the matrix to be updated/queried is a power of two. Each node of the quadtree built for handling operations on this matrix stores data for a square sub-matrix.\cite{QT3} Accordingly, the root node stores the data for the whole matrix and the leaves store the individual elements. Once the quadtree has been built, \textit{range-update range-query} operations can be performed by incorporating lazy propagation.

\vspace{2 mm}
\noindent
\textbf{Proposition 2.} For an $n \times n$ matrix, the worst case time complexity for query/update using a quadtree is $\Omega(n)$.

\vspace{1 mm}
\noindent
\textit{Proof.} We prove this by example. Consider a matrix $M$ with side-length $n$ ($n=2^k$). Quadtree recursively divides $M$ into square sub-matrices. Now, consider querying/updating a row in this matrix. No part of this row forms a square region except for the individual elements. This implies that no internal node in the quadtree stores data specifically for this row, or a part of it. Therefore, to update/query, we need to visit all the leaves associated with the elements of the row to be updated/queried. There are $n$ such leaves. Since we traverse each edge between the node and the leaves only once, the height of the tree is traversed only once. Thus, the time complexity of such an update/query becomes $\mathcal{O}(n)$. As there exists at least one operation with worst case time complexity $\mathcal{O}(n)$, the worst case running time of a quadtree is $\Omega(n)$.

\vspace{2 mm}
In his paper\cite[p. 240]{HS2}, Samet states that the worst-case memory requirement for building a quadtree occurs when the region concerned corresponds to checker-board pattern. We encounter this case when building quadtree over a 2D array. According to Samet, the number of nodes in such cases is a function of $r$, where $r$ is height (also known as resolution) of the quadtree. Since, each node in a quadtree has four children, therefore, the total number of nodes in a tree with height $r$ is $\frac{4^{r}-1}{3}$.\cite{QTN} Hence, the memory requirement is $\mathcal{O}(4^{r})$.

\vspace{2 mm}
For an $n \times n$ matrix, the height of the quadtree is $\log_2 n +1$ (height of a quadtree is same as the depth of recursion\cite[p. 2--3]{QTN}). Therefore, the memory requirement is $\mathcal{O}(4^{\log_2 n + 1})$. On simplifying this expression, we get $(2n)^2$.

The 3D version of the quadtree --- known as `octree' --- recursively divides a 3D space into octants. By similar analysis as for quadtree, we can say that an octree has worst case running time of $\Omega(n^2)$. The analysis of memory requirement can be done on the same lines as for quadtree, and is given by $\mathcal{O}(8^{\log_2 n + 1})$. This simplifies to $\mathcal{O}((2n)^3)$.

\vspace{2 mm}
In conclusion, for a quadtree generalized to $d$-dimensions, the worst-case time and space complexities are $\Omega(n^{d-1})$ and $\mathcal{O}((2n)^d)$ respectively.

\section{Binary Indexed Trees and associated algorithms}
\subsection{Description of the data structure}
Binary Indexed Tree (BIT) or Fenwick Tree\cite{BIT} is a data structure commonly used for calculating prefix sums efficiently. A BIT works by storing partial cumulative sums. For example, index 8 of BIT contains the cumulative sum of elements from 1 to 8, \textit{i.e.}, $rsum(1\, :\, 8)$. Similarly, index 6 stores $rsum(5\, :\, 6)$, and so on. For answering queries, BIT combines these stored partial sums. A segment tree can also be used to perform the functions of a BIT, but BITs are easier to code and have a lower constant of complexity. Hence, we base our algorithm on them. A BIT can handle \textit{range-update point-query} operations or \textit{range-update point-query} operations, but not both simultaneously. \textit{Range-update range-query} includes both the sets of operations. By this, we mean that \textit{range-update range-query} is a general case of both \textit{range-update point-query} and \textit{point-update range-query}.


\subsection{Point-update range-query on BIT}
This is the standard set of operations that a BIT can handle. As explained in the paper by Fenwick\cite{BIT}, a BIT is capable of updating the value at a particular point, and querying the cumulative sum up till a particular point. Throughout this paper, we use $updatep(bit,\, x,\, c)$ to denote a \textit{point-update} operation on a 1D BIT $bit$ which adds $c$ to element at position $x$. Similarly, we use $queryr(bit,\, x)$ to denote the operation which returns $rsum(1 :\, x)$, \textit{i.e.}, the cumulative sum of elements up till position $x$. Any arbitrary range $[x\, :\, y]$ can be queried by querying $[1\, :\, x-1]$ and $[1\, :\, y]$.

\vspace{2 mm}
We give pseudo-codes for both the functions. The algorithms were devised by Fenwick\cite{BIT}.

\vspace {3 mm}
\noindent
\captionof*{algorithm}{Point-update on 1D BIT}
\begin{algorithmic}[1]

\Function {updatep}{bit, x, c}
        \State $i \gets x$
        \While {$i \leq n$}
            \State $bit[i] \gets bit[i] + c$
            \State $i \gets i + bitAnd(i, -i)$
        \EndWhile
\EndFunction
\end{algorithmic}

\vspace {3 mm}
\noindent
\captionof*{algorithm}{Range-query on 1D BIT}
\begin{algorithmic}[1]

\Function {queryr}{bit, x}
        \State $sum \gets 0$
        \State $i \gets x$
        \While {$i > 0$}
            \State $sum \gets sum + bit[i]$
            \State $i \gets i - bitAnd(i, -i)$
        \EndWhile
        \State \Return $sum$
\EndFunction
\end{algorithmic}

\vspace{3 mm}
A 2D BIT uses a 2D array to store values. Just as each index in a 1D BIT stores the cumulative sum of a particular sub-array, similarly, each index in a 2D BIT stores the cumulative sum of a particular sub-matrix. For example, the index $(8, 8)$ stores the cumulative sum of all elements in the range $[1, 1\, :\, 8, 8]$, \textit{i.e.}, $rsum(1, 1\, :\, 8, 8)$. Similarly, index $(6, 4)$ stores $rsum(5, 1\, :\, 6, 4)$, and so on. It can be observed that a 2D BIT can be treated as a BIT of 1D BITs. Throughout this paper, we use $updatep(bit,\, (x, y),\, c)$ to denote a \textit{point-update} operation on 2D BIT named $bit$, which adds constant $c$ to the element at position $(x, y)$. Similarly, we use $queryr(bit,\, (x, y))$ to denote the operation which returns $rsum(1, 1\, : \,x, y)$. Further using the inclusion-exclusion principle, any sub-matrix can be queried.

\vspace{2 mm}
For clarity, we present pseudocodes for $updatep(bit,\, (x, y),\, c)$ and $queryr(bit,\, (x, y) )$ operations on a 2D BIT of side-length $n$:

\vspace {3 mm}
\noindent
\captionof*{algorithm}{Point-update on 2D BIT}
\begin{algorithmic}[1]

\Function {updatep}{bit, (x, y), c}
        \State $i \gets x$
        \While {$i \leq n$}
            \State $j \gets y$
            \While{$j \leq n$}
                \State $bit[i][j] \gets bit[i][j] + c$
                \State $j \gets j + bitAnd(j, -j)$
            \EndWhile
            \State $i \gets i + bitAnd(i, -i)$
        \EndWhile
\EndFunction
\end{algorithmic}

\vspace {3 mm}
\noindent
\captionof*{algorithm}{Range-query on 2D BIT}
\begin{algorithmic}[1]

\Function {queryr}{bit, (x, y)}
        \State $sum \gets 0$
        \State $i \gets x$
        \While {$i > 0$}
            \State $j \gets y$
            \While{$j > 0$}
                \State $sum \gets sum + bit[i][j]$
                \State $j \gets j - bitAnd(j, -j)$
            \EndWhile
            \State $i \gets i - bitAnd(i, -i)$
        \EndWhile
        \State \Return $sum$
\EndFunction
\end{algorithmic}

\vspace{3 mm}
The algorithms can be generalized to BITs of any number of dimensions on the same lines. The time complexity of the functions for a $d$-dimensional BIT is $\mathcal{O}(\log^d n)$.


\subsection{Range-update point-query on BIT}
We take an array $arr$. Each of its values is initially set to 0. We wish to add a constant $c$ to all the elements in the sub-array $arr[i\, :\, j]$. We want to do this multiple times for arbitrary $i$, $j$ and $c$. After some of the update operations, we want to know the value of an arbitrary element $arr[k]$. This is the simplest case of \textit{range-update point-query}. Using the algorithms and functions of \textit{point-update range-query}, a BIT can be made to handle such operations.

\vspace{2 mm}
Throughout this paper, we use $updater(bit,\, x_1,\, x_2,\, c)$ to denote an update operation on a 1D BIT $bit$, which adds constant $c$ to each element in the sub-array $[x_1\, :\, x_2]$. We use $queryp(bit,\, x)$ to denote the operation that returns the value of the element at position $x$. We first give the pseudocode for the functions, and then explain their working.

\vspace {3 mm}
\noindent
\captionof*{algorithm}{Range-update on 1D BIT}
\begin{algorithmic}[1]

\Function {updater}{bit, $x_1$, $x_2$, c}
        \State $updatep(x,\, c)$
        \State $updatep(y+1,\, -c)$
\EndFunction
\end{algorithmic}

\vspace {3 mm}
\noindent
\captionof*{algorithm}{Point-query on 1D BIT}
\begin{algorithmic}[1]

\Function {queryp}{bit, x)}
        \State $val \gets 0$
        \State $i \gets x$
        \While {$i > 0$}
            \State $val \gets sum + bit[i]$
            \State $i \gets i - bitAnd(i, -i)$
        \EndWhile
        \State \Return $val$
\EndFunction
\end{algorithmic}

\vspace{3 mm}
By using two \textit{point-update} operations, we can perform a \textit{range-update}. The algorithm for query operation remains the same as in the first set of operations. As can be noticed, the update algorithm works on the inclusion-exclusion principle. Whenever we have to update a range $[x_1\, :\, x_2]$ with a constant $c$, the value at position $x$ of the $bit$ is increased by $c$. Due to this increase, $rsum(1\, : \,x)$, where $x_1 \leq x$, increases by $c$. The value of $rsum(1\, : \,x)$, where $x > x_2$, does not increase since all the elements after $x_2$ are updated with $-c$. Thus, an $updater$ operation only increases the values of the elements in the specified range.\\
The query algorithm remains the same. However, since an update operation only increases the value of the elements in a specific range, and initially $bit$ is set to 0, thus, $queryp$ returns the sum of updates that affected position $x$, \textit{i.e.}, the updated value of element at $x$. 

\vspace{2 mm}
We extend this algorithm to two dimensions. Henceforth, we use $updater(bit,\, (x_1, y_1),\, (x_2, y_2),\, c)$ to denote an update operation on a 2D BIT $bit$, which adds constant $c$ to each element in the sub-array $[x_1, y_1\, :\, x_2, y_2]$. We use $queryp(bit,\, (x, y) )$ to denote the operation that returns the value of the element at position $x, y$.

\vspace{2 mm}
Below, we give the pseudocode for the functions.

\vspace {3 mm}
\noindent
\captionof*{algorithm}{Range-update on 2D BIT}
\begin{algorithmic}[1]

\Function {updater}{bit, $(x_1, y_1)$, $(x_2, y_2)$, c}
        \State $updatep(bit,\, (x_1, y_1),\, c)$
        \State $updatep(bit,\, (x_2+1, y_1),\, -c)$
        \State $updatep(bit,\, (x_1, y_2+1),\, -c)$
        \State $updatep(bit,\, (x_2+1, y_2+1),\, c)$
\EndFunction
\end{algorithmic}

\vspace {3 mm}
\noindent
\captionof*{algorithm}{Point-query on 2D BIT}
\begin{algorithmic}[1]

\Function {queryp}{bit, (x, y)}
        \State $val \gets 0$
        \State $i \gets x$
        \While {$i > 0$}
            \State $j \gets y$
            \While{$j > 0$}
                \State $val \gets sum + bit[i][j]$
                \State $j \gets j - bitAnd(j, -j)$
            \EndWhile
            \State $i \gets i - bitAnd(i, -i)$
        \EndWhile
        \State \Return $val$
\EndFunction
\end{algorithmic}

\vspace{3 mm}
The algorithms can be generalized to BITs of any number of dimensions on the same lines. The time complexity of the functions for a $d$-dimensional BIT is $\mathcal{O}(\log^d n)$.


\section{Proposed algorithm}
By using the functions of \textit{range-update point-query} described in section 7.3, we can make a BIT handle \textit{range-update range-query} operations. In this section, we put forward the algorithm for 1D version and 2D version, and then present a generalization for arbitrary number of dimensions.


\subsection{For 1D arrays}
An important observation to be made here, is that if $rsum(1 \,:\, x-1)$ and $rsum(1\, :\, y)$ can be computed efficiently, then $rsum(x\, :\, y)$ can be calculated in constant time. This is because,

\vspace{-5 mm}
\begin{align}
rsum(x\, :\, y) = rsum(1\, :\, y)-rsum(1\, :\, x-1)
\end{align}

We define an operation $query(i)$ which returns the value $rsum(1\, : \,i)$, and an operation $update(x,\, y,\, c)$ which updates the sub-array $[x\, :\, y]$ with constant $c$.

\subsubsection{Range-update}
Let there be an array $arr$, indexed from 1 to n, with all elements initially set to 0. At this point, $query(i)$ returns 0 for any index $i$. An update operation $update(3,\, 5,\, 4)$ is performed.\\
Now, $query(x)$ is called for an arbitrary index $x$. There are three possible cases:
\begin{enumerate}[i.]
\vspace{-2 mm}
\item $i < 3$
\item $3 \leq i \leq 5$
\item $i > 5$
\end{enumerate}

\vspace{1 mm}
\noindent
For the first case, the value returned by the function $query(i)$ is $0$.\\
For the second case, the value returned should be 4 if $i=3$, 8 if $i=4$ and 12 if $i=5$.\\
For the third case, $query(i)$ should return 12 for all $i$.

\vspace{2 mm}
From the example above, the following inferences can be made:\\
When an operation $update(x_1,\, x_2,\, c)$ is performed, there is no change in the value of $rsum(1\, : \,x)$ if  $x < x_1$. For indexes $x$, where $x_1 \leq x \leq x_2$, the value of $rsum(1\, : \,x)$ changes by $(c*x\,-\,c*(x_1-1))$. For $x > x_2$, $rsum(1\, : \,x)$ changes by $c*(x_2-x_1+1)$.

\vspace{2 mm}
Thus, after a \textit{range-update} operation $update(x_1,\, x_2,\, c)$, the change in the value of $rsum(1\, : \,x)$ varies with index $x$ linearly. For indexes which are less that $x$, there is no change. For an index $x$, which is between $x_1$ and $x_2$, the change is given by the function $(c*x\,-\,c*(x_1-1))$. For all indexes $x > x_2$, the change is given by the function $(0*x\,+\,c*(x_2-x_1+1))$.

\vspace{2 mm}
As linear functions can be added, if the sum of all the functions (referred to as `net function') for an index $x$ is known, the total change to the initial value of $rsum(1\, : \,x)$ can be calculated by putting $x$ into the net function. To define a linear function at an index, we need to know the coefficient of the variable $x$ (henceforth referred to as `co-efficient'), and the value of the term independent of $x$ (henceforth referred to as `independent term'). The variable term and the independent terms in addition form the required linear function.

\vspace{2 mm}
If the sum of co-efficients and the sum of independent terms at an index are known, then the net function is also known. Hence, we keep two BITs. One BIT stores the co-efficients and the other stores the independent term. We name these BITs $bitc$ and $biti$ respectively. Whenever a function $ax + b$ (where $x$ is the index) is to be added to a range $[x_1 : x_2]$, we update the range $[x_1\, :\, x_2]$ of $bitc$ with $a$, and the range $[x_1\, :\, x_2]$ of $biti$ with $b$. When we need to calculate $rsum(1\, : \,x)$, for some $x$, we can perform \textit{point-query} operation at position $x$ in both the BITs to get the co-efficient and independent term of the net function.
Thus, by performing \textit{range-update point-query} on two BITs, we can keep track of the net function for each index.

\vspace{2 mm}
For an operation $update(x_1,\, x_2,\, c)$, the functions which need to be added, along with the ranges, are as follows.

\begin{enumerate}
\item Function $cx - c(x_1-1)$ to all indexes $x$ such that $x_1 \leq x \leq x_2$.
\item Function $c(x_2-x_1+1)$ to all indexes $x$ such that $x_2 < x \leq n$.
\end{enumerate}

Below, we present the pseudocode of the $update$ function. Note that we have used the $updater$ function from section 7.3 to perform the required \textit{range-updates} on the two BITs.

\vspace {3 mm}
\noindent
\captionof*{algorithm}{Range-update on 1D BIT}
\begin{algorithmic}[1]

\Function {update}{$x_1$, $x_2$, c}
        \State $updater(bitc,\, x_1,\, x_2,\, c)$
        \State $updater(biti,\, x_1,\, x_2,\, -c(x_1-1))$
        \State $updater(biti,\, x_2+1,\, n,\, c(x_2-x_1+1))$
\EndFunction
\end{algorithmic}

\vspace{3 mm}
The complexity of the \textit{range-update} operation is $\mathcal{O}(\log n)$.

\subsubsection{Range-query}
Whenever the value of $rsum(1\, : \,x)$, for an arbitrary index $x$, is needed, we call $queryp(bitc,\, x)$ on $bitc$ to get the coefficient, and on $queryp(biti,\, x)$ to get independent term of the net function. Multiplying the value obtained from $bitc$ with $x$, and adding the value obtained from $biti$ gives the net change in the value of $rsum(1\, : \,x)$. Since, we assumed all values to be 0 initially, hence, we get the actual value $rsum(1\, : \,x)$.

\vspace{2 mm}
Below, we present the pseudocode of the $query$ function. Note that we have used the $queryp$ function from section 7.3 to get the values of co-efficients and independent terms at the required positions in the BITs.

\vspace {3 mm}
\noindent
\captionof*{algorithm}{Range-query on 1D BIT}
\begin{algorithmic}[1]

\Function {query}{x}
        \State $a \gets queryp(bitc,\, x)$
        \State $b \gets queryp(biti,\, x)$
        \State return $a*x + b$
\EndFunction
\end{algorithmic}

\vspace{3 mm}
The complexity of the \textit{range-query} operation is $2*O(\log n)$, since 2 BITs are involved.


\subsection{Extension to 2D arrays}
We extend the algorithm devised above to matrices (2D arrays). An important observation based on inclusion-exclusion principle is that if $rsum(1, 1\, : \,x,y)$ can be computed efficiently for arbitrary $x$ and $y$, then $rsum(x_1, y_1 \,: \,x_2, y_2)$, for arbitrary $x_1$, $y_1$, $x_2$, $y_2$, can be computed efficiently too. This is because,

\vspace{-5 mm}
\begin{align}
\begin{split}
    rsum(x_1, y_1 : x_2, y_2) =\,& rsum(1, 1 : x_2, y_2)-rsum(1, 1 : x_2, y_1-1)\\
         & -rsum(1, 1 : x_1-1, y_2)+rsum(1, 1 : x_1-1, y_1-1)
\end{split}
\end{align}

We assume an operation $query( (x, y) )$ which returns the value of $rsum(1, 1\, : \,x,y)$, and an operation $update( (x_1, y_1),\, (x_2, y_2),\, c)$ which adds $c$ to each element of the sub-matrix $[x_1, y_1 \,:\, x_2, y_2]$. Further, any range can be queried by using the inclusion-exclusion principle.


\subsubsection{Range-update}
We begin by analyzing the change in values of $rsum(1, 1\, : \,x,y)$, for arbitrary $(x, y)$, after an $update( (x_1, y_1),\, (x_2, y_2),\, c)$ is performed.

\begin{enumerate}
\item For all $(x, y)$ where $x1 \leq x \leq x2$ and $y1 \leq y \leq y2$, change in $rsum(1, 1\, : \,x,y)$ is given by $cxy - c(y_1-1)x - c(x_1-1)y + c$.

\item For all $(x, y)$ where $x2 < x \leq n$ and $y1 \leq y \leq y2$, change in $rsum(1, 1\, : \,x,y)$ is given by $c(x_2-x_1+1)y - c(y_1-1)(x_2-x_1+1)$.

\item For all $(x, y)$ where $x1 \leq x \leq x2$ and $y2 < y \leq n$, change in $rsum(1, 1\, : \,x,y)$ is given by $c(y_2-y_1+1)x - c(x_1-1)(y_2-y_1+1)$.

\item For all $(x, y)$ where $x2 < x \leq n$ and $y2 < y \leq n$, change in $rsum(1, 1\, : \,x,y)$ is given by $c(x_2-x_1+1)(y_2-y_1+1)$.
\end{enumerate}

We have to maintain functions with 4 terms, \textit{i.e.}, $xy$, $y$, $x$ and an independent term. Thus, we maintain 4 2D BITs, namely $bitxy$, $bitx$, $bity$ and $biti$. We use the function $updater$, devised in section 7.3 for performing the required \textit{range-update} operations in order to maintain the functions.

\vspace{2 mm}
Below we give the pseudocode for the $update$ function for 2D case.

\vspace {3 mm}
\noindent
\captionof*{algorithm}{Range-update on 2D BIT}
\begin{algorithmic}[1]

\Function {update}{($x_1$, $y_1$), ($x_2$, $y_2$), c}
        \State $updater(bitxy,\, (x_1, y_1),\, (x_2,y_2),\, c)$
        
        \State $updater(bitx,\, (x_1, y_1),\, (x_2, y_2),\, -c(y_1-1))$
        \State $updater(bitx,\, (x_1, y_2+1),\, (x_2, n),\, c(y_2-y_1+1))$
        
        \State $updater(bity,\, (x_1, y_1),\, (x_2, y_2),\, -c(x_1-1))$
        \State $updater(bity,\, (x_2+1, y_1),\, (n, y_2),\, c(x_2-x_1+1))$
        
        \State $updater(biti,\, (x_1, y_1),\, (x_2,y_2),\, c)$
        \State $updater(biti,\, (x_2+1, y_1),\, (n, y_2),\, -c(y_1-1)(x_2-x_1+1))$
        \State $updater(biti,\, (x_1, y_2+1),\, (x_2, n),\, -c(x_1-1)(y_2-y_1+1))$
        \State $updater(biti,\, (x_2+1, y_2+1),\, (n, n),\, c(x_2-x_1+1)(y_2-y_1+1))$
\EndFunction
\end{algorithmic}

\vspace{3 mm}
Since, there are 4 2D BITs and operation on each 2D BIT takes $\mathcal{O}(\log^2 n)$, the net complexity per update is given by $4*O(\log^2 n)$.


\subsubsection{Range-query}
In order to query the net change in the value fo $rsum(1, 1\,: \,x, y)$, we need to know the net function at $(x, y)$. We use the $queryp$ function devised in section 7.3 to perform the point queries at $(x, y)$.

\vspace{2 mm}
We present the pseudocode for the $query$ operation below.

\vspace {3 mm}
\noindent
\captionof*{algorithm}{Range-query on 2D BIT}
\begin{algorithmic}[1]

\Function {query}{(x, y)}
        \State $a \gets queryp(bitxy,\, (x, y))$
        \State $b \gets queryp(bitx,\, (x, y))$
        \State $c \gets queryp(bity,\, (x, y))$
        \State $d \gets queryp(biti,\, (x, y))$
        \State return $a*x*y + b*x + c*y + d$
\EndFunction
\end{algorithmic}

\vspace{3 mm}
Since, there are 4 2D BITs, the net complexity per query is given by $4*O(\log^2 n)$.


\subsection{Generalization to higher dimensions}
It is evident from the algorithm's extension to matrices that it can be generalized to any number of dimensions. The algorithm for the $d$-dimensional version can be constructed from the one for ($d-1$) dimensions.

\vspace{2 mm}
\noindent
\textbf{Proposition 3.} For handling \textit{range-update range-query} operations on a $d$-dimensional array, $2^d$ $d$-dimen\-sional BITs are required.

\vspace{1 mm}
\noindent
\textit{Proof.} We prove this by mathematical induction. We have shown that 2 BITs are required for handling \textit{range-update range-query} operations on a 1D array. We assume that $2^k$ BITs are required to handle operations on a $k$-dimensional array, for some whole number $k > 1$.
We now take a $(k+1)$-dimensional array $arr$. The array $arr$ can be understood as a 1D array, each of whose elements is a $k$-dimensional array. As per our assumption, for updating an element of $arr$, we further need $2^k$ $k$-dimensional BITs per element (since each element is a $k$-dimensional array). Having a $k$ dimensional BIT for each element is same as having one $(k+1)$-dimensional BIT. This implies that we need $2^k$ $(k+1)$-dimensional BITs to perform update an element of $arr$. As shown previously, we need two BITs to perform \textit{range-update range-query} operations on a 1D array. As a result, in total, we need $2^{k+1}$ $(k+1)$-dimensional BITs for $arr$.\\
By principle of mathematical induction, since the assertion is true for 1, and also true for some whole number $k+1$ whenever its true for $k$, therefore, its true for the entire set of whole numbers. Therefore, we need $2^d$ $d$-dimensional BITs to perform \textit{range-update range-query} on a $d$-dimensional array.\\
Alternatively, the conclusion can also be reached by realizing that a multi-linear function with $d$ variables has $2^d$ coefficients. Therefore, $2^d$ $d$-dimensional BITs are required.

\vspace{2 mm}
An update or query on a $d$-dimensional BIT, with $n$ elements along each dimension, requires $\mathcal{O}(\log^d n)$ time. By \textit{Proposition 3}, it is known that $2^d$ such BITs are required to handle operations. Therefore, the overall time complexity for each update/query operation is  $\mathcal{O}(4^d*\log^d n)$.


\section{Experimental comparison of running times}
In this section, we provide an experimental comparison between the execution times of the old and new algorithms. In the table below, we give the total time taken (in milliseconds) by the two algorithms to perform $100,000$ update/query operations on 2D and 3D arrays of different side-lengths ($n$).

\vspace{1 mm}
\begin{center}
\begin{tabular}{|c|c|c|c|c|}
\hline
\multirow{2}{*}{$\bm{n}$} & \multicolumn{2}{ |c| }{\textbf{Old Algorithm}} & \multicolumn{2}{ |c| }{\textbf{New Algorithm}}\\ \cline{2-5}
& \textbf{2D (Quadtree)} & \textbf{3D (Octree)} & \textbf{2D} & \textbf{3D}\\
\hline
10 & 81 & 522 & 93 & 693 \\
50 & 680 & 53051 & 238 & 1923 \\
100 & 888 & 219512 & 309 & 3186 \\
150 & 1387 & 488673 & 400 & 3992 \\
200 & 3184 & 910796 & 431 & 6357 \\
500 & 4420 & \- & 393 & \- \\
1000 & 8996 & \- & 525 & \- \\
4000 & 39126 & \- & 1644 & \- \\
\hline
\end{tabular}
\end{center}

\vspace{-3 mm}
\begin{center} \small{Table 1: Comparison of running times} \end{center}

\vspace{2 mm}
For the purpose of this experiment, we used tree-based implementation of quadtree and octree. The parameters for updates and query operations --- the sub-arrays to be updated/queried and the constants for the update operations --- were produced using the $rand()$ function in the \textit{stdlib.h} header file of \textit{C Library}. The parameters that we used were taken uniformly at random in the space of all parameters.

\vspace{2 mm}
\noindent
The implementations were done on a machine with Ubuntu 13.10 64-bit as the operating system, 8 GB RAM and Intel Core i3 3.1 GHz Sandy Bridge processor.

\vspace{2 mm}
Below, we plot the graph for number of instructions ($T$) vs. number of dimensions ($d$) for arrays of side-lengths ($n$) equal to 100, 1000 and 10000.

\begin{figure}[h]
\centering
\subfloat[$n=100$]{\includegraphics[width=120mm, height=60mm]{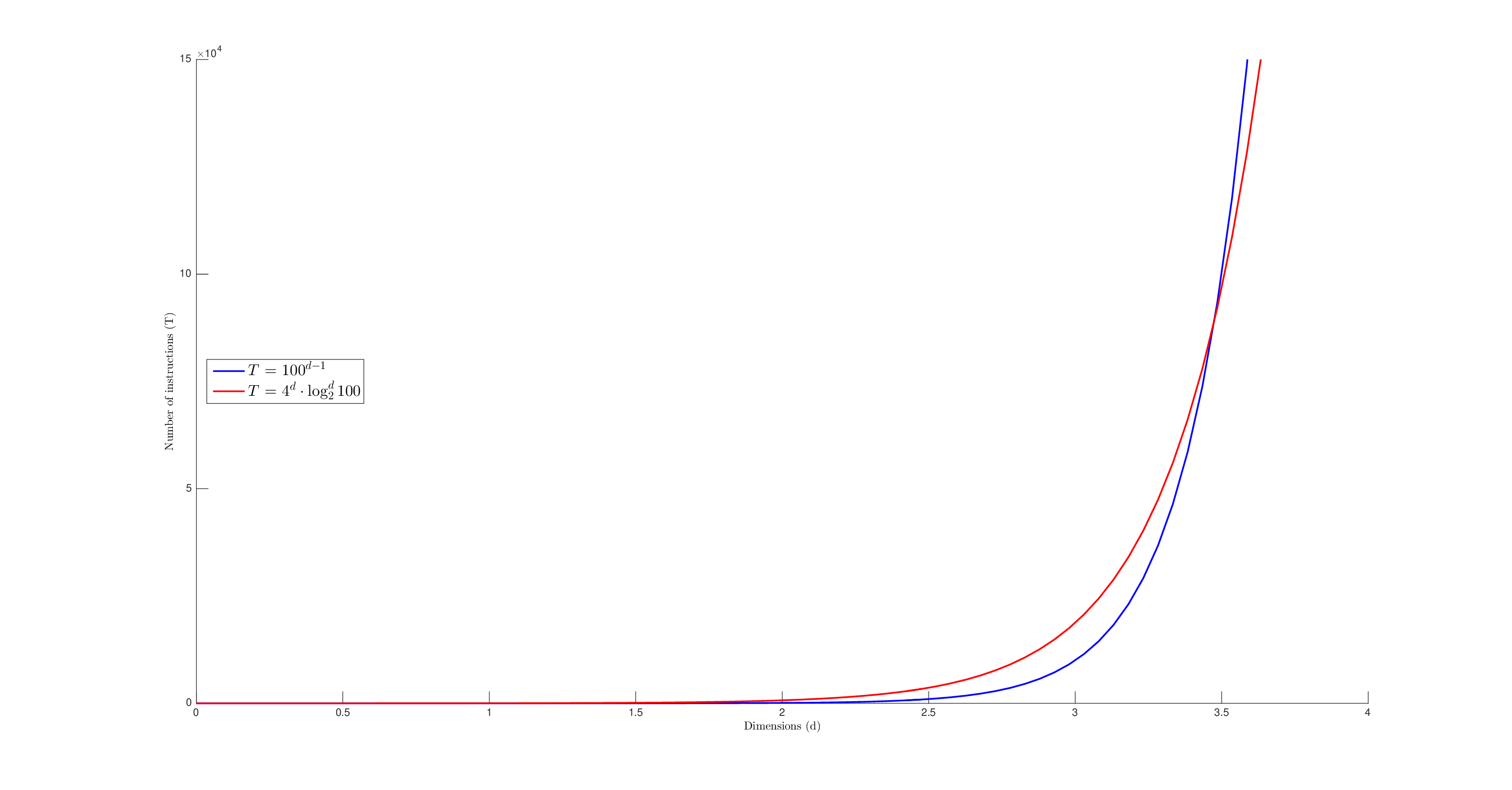}}
\end{figure}
\pagebreak
\begin{figure}[h]
\centering
\subfloat[$n=1000$]{\includegraphics[width=120mm, height=60mm]{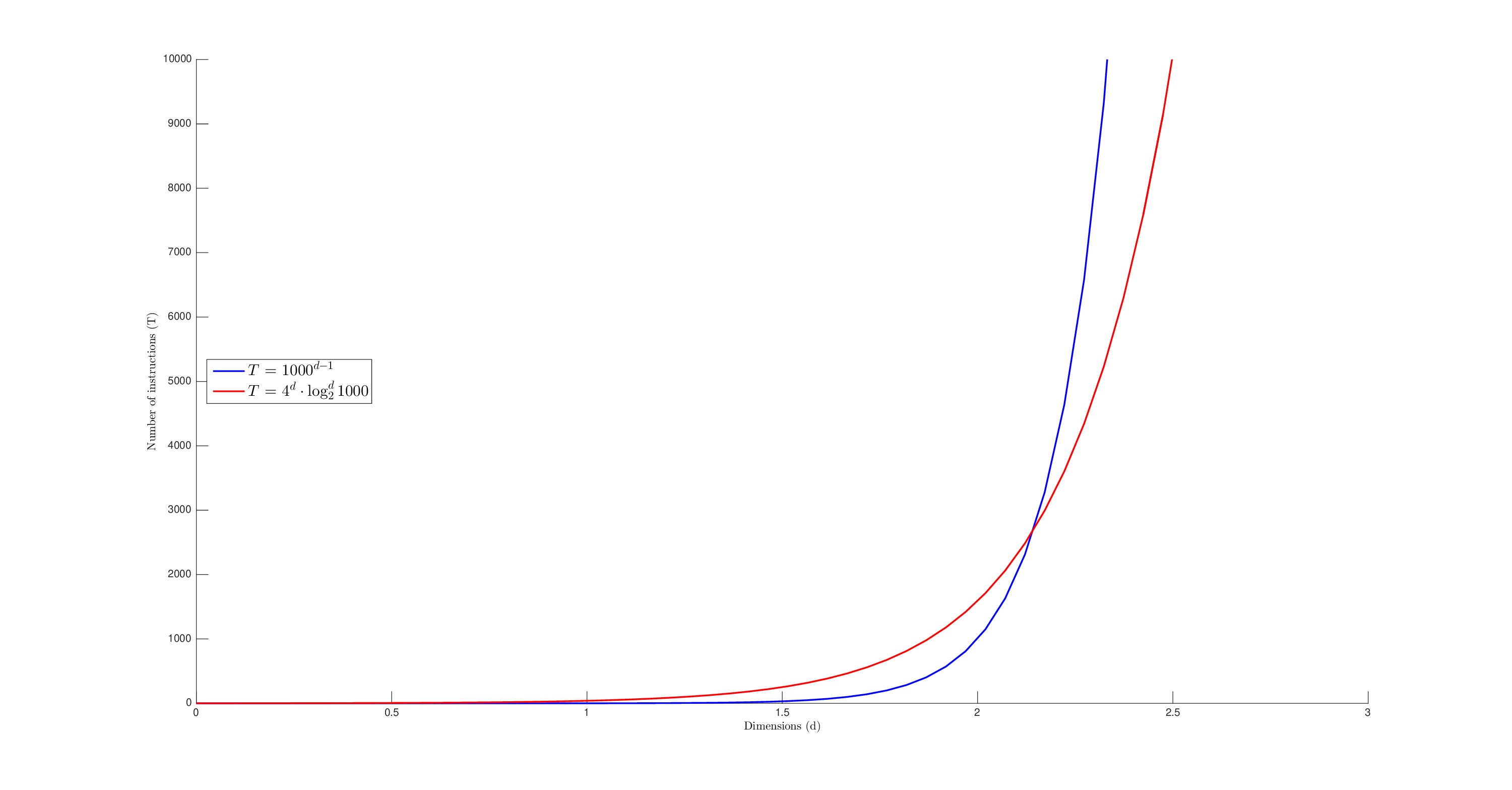}}\\
\end{figure}
\begin{figure}[h]
\centering
\subfloat[$n=10000$]{\includegraphics[width=120mm, height=60mm]{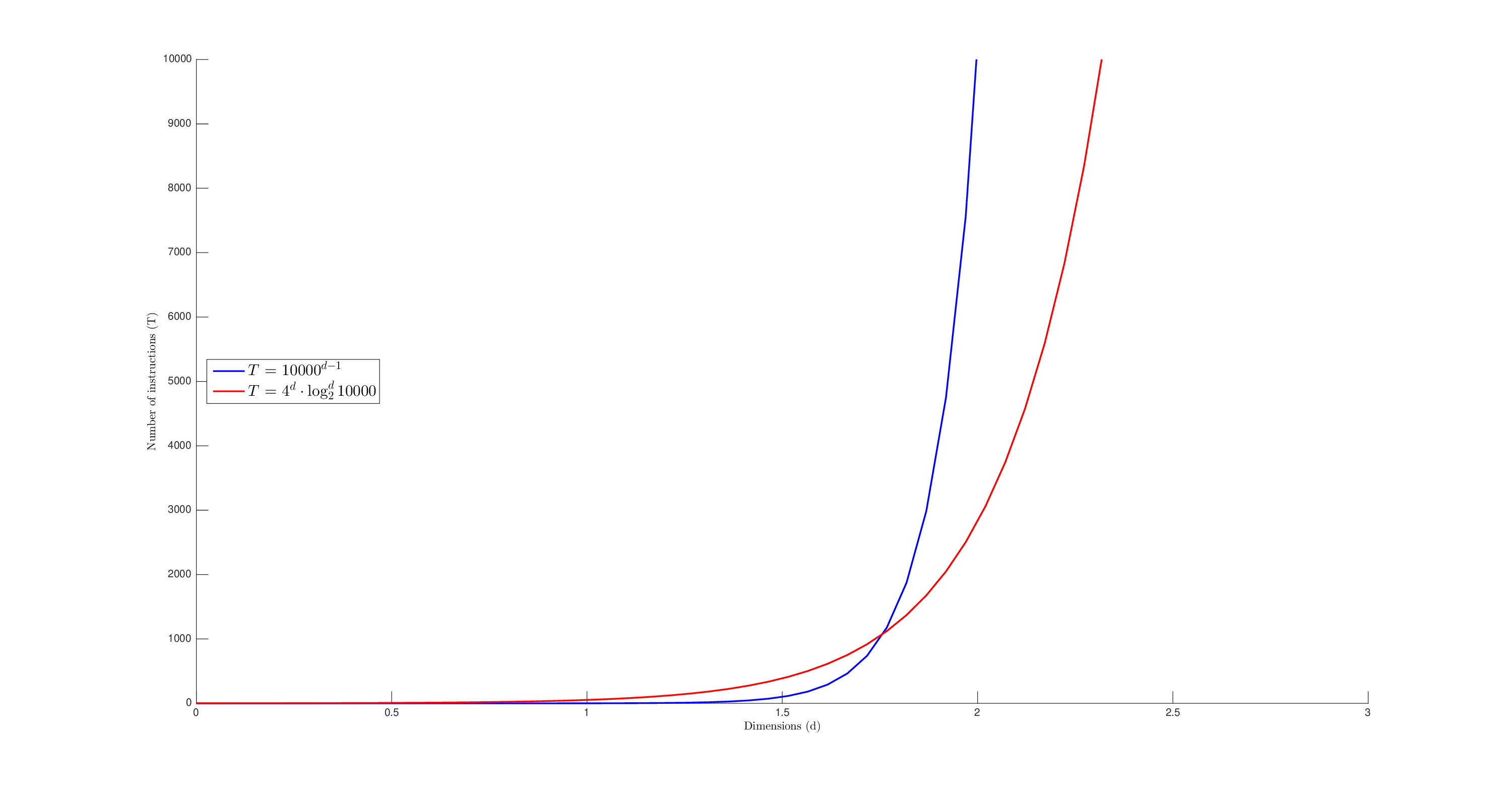}}\\
\caption{Number of instructions ($T$) vs. Number of dimensions ($d$)}
\end{figure}

\noindent
From the graphs and the table of experimental running times, the following inferences can be made.

\begin{itemize}
\item The graph for the old algorithm rises more steeply as compared to the graph for the new algorithm.
\item The new algorithm outperforms the old algorithm in the number of instructions per second and, consequently, in time of execution.
\item Efficiency of the new algorithm gets more pronounced as $n$ increases.
\end{itemize}


\section{Conclusion and future scope}
The algorithm that this paper proposes significantly reduces the time required for handling on-line \textit{range-update range-query} operations on sub-arrays of multidimensional arrays. However, the memory needed to execute the algorithm --- as in the case with previously known algorithms --- increases exponentially with the number of dimensions. We still believe that our algorithm is of practical significance since the number of dimensions rarely exceeds three in real world applications.



\end{document}